\documentclass[11pt,preprint]{aastex}
\usepackage{apjfonts}
\usepackage{placeins}

\shorttitle{ACCRETION OUTBURSTS} \shortauthors{Bae et al.}

\begin{document}

\title{Variable Accretion Outbursts in Protostellar Evolution}

\author{Jaehan Bae\altaffilmark{1},
Lee Hartmann\altaffilmark{1},
Zhaohuan Zhu\altaffilmark{2},
Charles Gammie \altaffilmark{3,4}}

\altaffiltext{1}{Dept. of Astronomy, University of Michigan, 500
Church St., Ann Arbor, MI 48105} 
\altaffiltext{2}{Department of Astrophysical Sciences, Princeton University,
4 Ivy Lane, Peyton Hall, Princeton, NJ 08544}
\altaffiltext{3}{Dept. of
Astronomy, University of Illinois Urbana-Champaign, 1002 W. Green
St., Urbana, IL 61801}
\altaffiltext{4}{Dept. of Physics, University
of Illinois Urbana-Champaign}

\email{jaehbae@umich.edu, lhartm@umich.edu, zhuzh@astro.princeton.edu, gammie@illinois.edu}

\newcommand\msun{M_{\odot}}
\newcommand\lsun{L_{\odot}}
\newcommand\msunyr{M_{\odot}\,{\rm yr}^{-1}}
\newcommand\be{\begin{equation}}
\newcommand\en{\end{equation}}
\newcommand\cm{\rm cm}
\newcommand\kms{\rm{\, km \, s^{-1}}}
\newcommand\K{\rm K}
\newcommand\etal{{et al}.\ }
\newcommand\sd{\partial}
\newcommand\mdot{\dot{M}}
\newcommand\rsun{R_{\odot}}
\newcommand\yr{\rm yr}

\begin{abstract}
We extend the one-dimensional, two-zone models of long-term protostellar disk evolution
with infall of Zhu et al. to consider the potential effects of a finite viscosity
in regions where the ionization is too low for the magnetorotational instability (MRI) to
operate (the ``dead zone'').  We find that the presence of a small but finite dead zone 
viscosity, as suggested by simulations of stratified disks with MRI-active outer layers,
can trigger inside-out bursts of accretion, starting at or near the inner edge of the
disk, instead of the previously-found outside-in bursts with zero dead zone viscosity,
which originate at a few AU in radius.  These inside-out bursts of accretion
bear a qualitative resemblance to the outburst behavior of one FU Ori object, V1515 Cyg,
in contrast to the outside-in burst models which more closely resemble the accretion events
in FU Ori and V1057 Cyg. 
Our results suggest that the type and frequency of outbursts are potentially a probe of
transport efficiency in the dead zone.
Simulations must treat the inner disk regions,
$R \lesssim 0.5$~AU, to show the detailed time evolution of accretion outbursts in general
and to observe the inside-out bursts in particular. 
\end{abstract}


\section{Introduction}
The (re)discovery of the magnetorotational instability \citep[MRI: e.g.,][and references
therein]{bh98},
appears to resolve the long-standing problem of the anomalous
viscosity in sufficiently ionized accretion disks.  To a crude approximation, this
validates the use of \citet{ss73} ``$\alpha$ viscosity" disks, though there are
differences in detail \citep[e.g.,][]{bp99, gammie96}.  Constant $\alpha$ disks have
been employed in many situations, including the evolution of
pre-main sequence disks \citep[e.g.,][]{hartmann98}.
However, as pointed out by \citet{gammie96}, thermal ionization levels 
in protostellar and protoplanetary disks generally are so low that 
it is unlikely that the MRI operates everywhere.
Gammie suggested that transport in these regions might be limited to surface
``active'' layers, in which non-thermal ionization (cosmic rays, stellar X-rays) could allow the
MRI to operate, while in the central regions of the disk there could be a non-viscous ``dead zone''.  
A rich variety of phenomena are thus enabled beyond simple (quasi-steady)
viscous disks \citep[][]{vb05, vb06, vb07, vb09, z10a, z10b, martin11, martin12a, martin12b}.

\citet{gammie96} noted that if the MRI-active layer has a roughly constant surface density,
the result is a pileup of material in the inner disk which eventually could become
gravitationally unstable and result in a rapid burst of accretion, perhaps producing
an FU Ori outburst \citep{hk96}.  While early models of FU Ori accretion
events relied on traditional thermal instability theory 
\citep{clarke90,bell94},
\citet{zhu07} showed that the rapidly-accreting region of FU Ori extends much further in radius
from the central star than can be achieved with this theory.  The very large amount of
mass accreted in one of FU Ori's outbursts ($\sim 10^{-2}~\msun$) over such short timescales
($\sim 10^2$~yr) requires a large amount of material to be present in the disk at relatively
small radii.  This is a natural result of models in which gravitational instability (GI) triggers
the outbursts \citep[][]{armitage01, vb06, vb07, vb09, zhu09c, z10a, z10b, 
martin11, martin12a, martin12b}.

\citet{z10b} presented one-dimensional, two-layer evolutionary disk models including
infall from a rotating protostellar cloud, and showed that they could qualitatively reproduce the main
features of the outbursts of FU Ori and V1057 Cyg, with rapid rise times and slowly-decaying
accretion.  The simulations included irradiation from the central star, but did not
take into account the accretion luminosity, which during outbursts can be far larger than
the stellar photospheric radiation.  While the geometry of disk accretion may not favor
self-irradiation \citep{bell99}, during infall the dusty opaque envelope will act as
a blanket, reradiating a significant portion of the accretion luminosity toward the disk.
In addition the \citet{z10b} calculations assumed that the central layer - the dead zone - 
had no viscosity unless it became gravitationally unstable.
However, detailed shearing box simulations of the MRI in disks with a stratified structure 
and a resistivity that increases toward the midplane indicate that MHD turbulence 
generated in the upper MRI-active layers produces some hydrodynamic turbulence in the 
inactive layers \citep{flemingstone03, okuzumi11, gressel12}.
Furthermore, these investigations suggest that this turbulence creates Maxwell stresses,
which result in a small but non-zero viscosity that can transport angular momentum outward and thus mass inward.  

These considerations motivate further investigations of protostellar accretion using the
two-layer one-dimensional model.  We find that the position and properties of the inner
boundary, the inclusion of irradiation by the accretion luminosity generated in the inner disk,
and any non-zero dead zone viscosity have significant effects on the resulting
bursts of mass accretion.  Our treatment is sufficiently limited to preclude detailed
predictions, but the qualitative behavior is suggestive, given that continuing
observations of protostars and pre-main sequence stars are increasingly found to exhibit
a wide variety of accretion events, beyond the large FU Ori outbursts
\citep{hk96, herbig08, muzerolle05, reipurth10, aspin10, covey11, lorenzetti12}.

\section{Methods}

We use a modified version of the one-dimensional, two-zone disk model 
previously introduced in \citet{z10a,z10b}, including changes to
the infall model, enhanced disk heating, and viscosity in the dead zone.
Here we review our scheme.

\subsection{Surface density evolution}
\label{sec:density}

The surface density of a disk is evolved 
based on the mass and angular momentum conservation equations in cylindrical coordinates,
\be\label{eqn:mass}
2\pi R {\partial \Sigma_i \over \partial t} - {\partial \dot{M}_i \over \partial R} = 2\pi g_i (R, t)
\en
and
\be\label{eqn:ang_momentum}
2\pi R {\partial\over \partial t}(\Sigma_i R^2 \Omega) - {\partial \over \partial R}(\dot{M}_i R^2 \Omega) = 2\pi {\partial \over \partial R} (R^2 W_{R\phi, i}) + 2\pi \Lambda_i (R, t),
\en
where $\Sigma_i$ is the surface density, $\Omega$ is the angular frequency, $\dot{M}_i$ is the radial mass flux, $W_{R\phi, i}=R\Sigma_i \nu_i d\Omega/dR$, and $\nu_i$ is the viscosity.
The subscript $i$ denotes either the active layer (``$a$") or dead zone (``$d$").
The terms $2\pi g_i (R, t)$ and $2\pi \Lambda_i (R, t)$ are the mass and angular momentum flux per unit distance of infall material from an envelope cloud \citep{Cassen81}.
Then, assuming instantaneous centrifugal balance (see \citealt{z10b}), 
equations (\ref{eqn:mass}) and (\ref{eqn:ang_momentum}) can be simplified to 
\be\label{eqn:mdot}
\dot{M}_i = 6\pi R^{1/2} {\partial \over \partial R} (R^{1/2} \Sigma_{i} \nu_i) + {2\pi R^2 \Sigma_i \over M_R} {\partial M_R \over \partial t} - 4\pi \left( {R \over GM_R}\right)^{1/2} (\Lambda_i(R,t)-g_i(R, t) R^2 \Omega(R)),
\en
where $M_R$ is the sum of the mass of the central star and the disk mass within $R$.
We then sequentially solve Equations (\ref{eqn:mdot}) and (\ref{eqn:mass}) to evolve the disk.

In \citet{z10b} mass and angular momentum were added to the
disk using the infall model of \citet{Cassen81},
\be
g(R,t) = {\dot{M}_{\rm in} \over {4\pi R_c}} \left( 1- {R \over R_c} \right)^{-1/2} {\rm if}~R \le R_c,
\en
\be
g(R,t) = 0~{\rm if}~R > R_c,
\en
and
\be
\Lambda(R,t) = g(R, t) R \left( {GM_c \over R_c} \right)^{1/2}~{\rm if}~R \le R_c,  
\en 
\be
\Lambda(R,t) = 0~{\rm if}~R > R_c.
\en  
Here $R_c$ is the centrifugal radius, that is the outer radius at which mass is added
to the disk at time $t$, and $\dot{M}_{\rm in}=0.975c_s^3/G$ is a constant total infall mass rate at a given cloud temperature \citep{shu77,terebey84}.
However, this model has a singularity at $R=R_c$; 
with finite grids this potentially causes non-convergent behavior at different grid
resolutions.  To avoid this we modified the infall model to 
eliminate the singularity by using a constant mass flux per unit distance. 

Another problem occurs at the inner boundary, which is difficult to make very small because
for numerical reasons (short time steps, dust evaporation, etc.) as well as fundamental uncertainties;
for example, is the disk truncated by a stellar magnetosphere, and if so where, and 
is there an outflow from the inner disk edge.  In addition,
mass infall right at the inner boundary produces different results depending upon
on precisely which inner boundary radius we choose.
To minimize these problems we take inner boundary radii which are relatively small but still
comfortably outside the expected point of magnetospheric truncation.  We further
assume that envelope material does not fall onto the disk inside 
$0.2R_c$, justified on the basis that the well-known emergence of jets and outflows
seen in even the earliest protostellar phases should prevent the lowest-angular momentum
material from reaching the disk or star \citep{reipurth01}.  By tying the inner radius of
infall to $R_c$ we effectively assume that the same streamline denotes the boundary
between outflow and inflow, such that the outflow cone retains the same opening angle,
in this case, a half-angle of $26.6^{\circ}$.  This particular choice of opening
angle is arbitrary and adopted mainly for numerical convenience. 

\begin{figure}
\begin{center}
\includegraphics{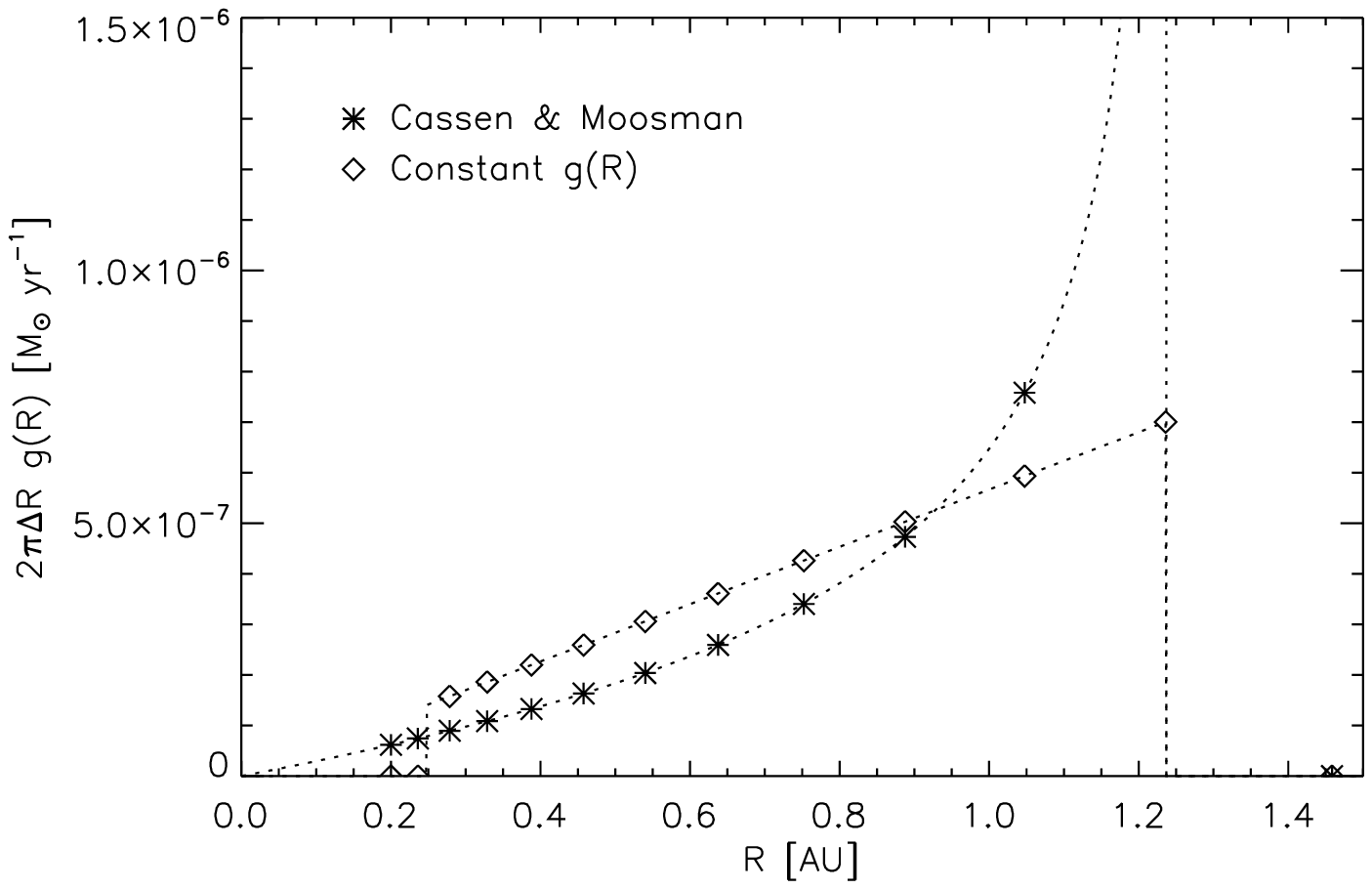}
\caption{The mass infall rate at $t=0.05$~Myr of \citet{Cassen81} model (asterisks) and our constant $g(R)$ model (diamonds).
The centrifugal radius at the time is $R_c\sim1.24$~AU.
\citet{Cassen81} model gives $\sim1.3\times10^{-5}~M_{\odot}~{\rm yr}^{-1}$ of mass infall rate at the nearest grid to $R_c$, which is about an order of magnitude large to be fitted in the figure. 
The mass infall rate inside $0.2~R_{\rm c}$ is set to zero in our model to imitate protostellar outflows (see text).
Dotted curves show analytic estimates of the mass infall rate of each model.}
\end{center}
\label{fig:rc}
\end{figure}

The mass infall rate of the modified model is 
\be
g(R,t) = {\dot{M}_{\rm in} \over {2\pi R_c}}~{\rm if}~0.2R_c \le R \le R_c
\en
and
\be
g(R,t) = 0~{\rm if}~R<0.2R_c~{\rm or}~R> R_c.
\en
The corresponding angular momentum added to the disk per unit distance is 
\be
\Lambda(R,t) = g(R, t) R \left( {GM_c \over R_c} \right)^{1/2} ~{\rm if }~0.2R_c \le R \le R_c
\en
and
\be
\Lambda(R,t) = 0~{\rm if}~R<0.2R_c~{\rm or}~R> R_c.
\en
A comparison of the radial mass infall profile of our model to
that of the \citet{Cassen81} model is presented in Figure \ref{fig:rc}.
In total, our model adds $11~\%$ less angular momentum to the disk 
per unit mass infall than the \citet{Cassen81} model.
These modifications result in better convergence with increasing grid resolution.

\subsection{Temperature evolution}

The disk layer temperatures are determined by the 
balance between heating and radiative cooling.
For the active layer, the energy equation is 
\begin{eqnarray}
\label{eqn:energy_a}
C_{\Sigma, a} \partial_t T_a & = & Q_{{\rm heat},a} - Q_{{\rm cool},a}
\nonumber\\
& = & Q_{{\rm vis},a} + Q_{{\rm infall},a} + Q_{{\rm grav},a} + {16 \over 3} \sigma \left( T_{\rm ext}^4 {\tau_a \over 1 + \tau_a^2} + T_d^4 {\tau_d \over 1 + \tau_d^2} \right)
\nonumber\\
& - &{16 \over 3}\sigma T_a^4 \left({\tau_a \over 1 + \tau_a^2} + {\tau_d \over 1 + \tau_d^2} \right) ,
\end{eqnarray}
where $C_{\Sigma, a} = \Sigma_a c_{s,a}^2 / T_a$ is the heat capacity of the active layer.
Here $T_{\rm ext}$ characterizes the heating flux due to the irradiation of the disk by the
stellar and (inner disk) accretion luminosity,
and $\tau_a$ and $\tau_d$ are the optical depths of the active layer and the dead zone,
respectively, 
\be
\tau_a = {1 \over 2} \Sigma_a \kappa (\rho_a, T_a)
\en
and 
\be
\tau_d = {1 \over 2} \Sigma_d \kappa (\rho_d, T_d)
\en
using the Rosseland mean opacity $\kappa$ taken from \citet{zhu09a}.
In Equation (\ref{eqn:energy_a}), the first three terms are local heating of the active layer due to the viscosity, the infall, and the gravitational potential energy change, respectively.
The fourth term consists of the external heating (see below) and the 
radiative heating from the underlying dead zone.
The last term includes the radiative cooling toward each side of the active layer.

The energy equation of the dead zone is similar to that of the active layer,
\be
C_{\Sigma, d} \partial_t T_d = Q_{{\rm heat},d} - Q_{{\rm cool},d},
\en
where $C_{\Sigma, d} = \Sigma_d c_{s,d}^2 / T_d$ is the heat capacity of the dead zone.
If the active layer is optically thick, the energy equation is
\be
\label{eqn:energy_d_thick}
C_{\Sigma, d} \partial_t T_d = Q_{{\rm vis},d} + Q_{{\rm infall},d} + Q_{{\rm grav},d} + {16 \over 3}\sigma T_a^4 {\tau_d \over 1 + \tau_d^2} - {16 \over 3}\sigma T_d^4 {\tau_d \over 1 + \tau_d^2}.
\en
On the other hand, if the active layer is optically thin, the incident flux from the outside of the dead zone would be $\sigma(\tau_a T_a^4 + T_{\rm ext}^4)$ so that the energy equation is
\be
\label{eqn:energy_d_thin}
C_{\Sigma, d} \partial_t T_d = Q_{{\rm vis},d} + Q_{{\rm infall},d} + Q_{{\rm grav},d} + {16 \over 3}\sigma (\tau_a T_a^4 + T_{\rm ext}^4) {\tau_d \over 1 + \tau_d^2}  - {16 \over 3}\sigma T_d^4 {\tau_d \over 1 + \tau_d^2}.
\en
Again, the first three terms in equations (\ref{eqn:energy_d_thick}) and (\ref{eqn:energy_d_thin}) represent local heating, while the last two terms account for radiative heating from the 
outside of the dead zone and the radiative cooling.

In the energy equations, the viscous heating is
\be
Q_{{\rm vis},i} = {3 \over 2} W_{R\phi,i}\Omega,
\en
where $W_{R\phi,i}=(3/2)\Sigma_i \nu_i \Omega$ and $\nu_i = \alpha_i c_{s,i}^2/\Omega$.
The viscosity parameter $\alpha_i$ is explained in detail in the next section. 

During infall the added material has smaller specific angular momentum than the disk material
at the same radius.  This results in a readjustment of the disk such a way that material moves inward.
As we are assuming effectively instantaneous centrifugal balance, the increase in the
gravitational potential energy driven by the readjustment process must be accompanied by the corresponding energy release.
Here we assume that this heats the active layer only ($Q_{{\rm infall},d}=0$), as this
is the material directly impacted by the infalling matter.
The heating by infalling material is then
\be
Q_{{\rm infall},a} = {GM_* \dot{M}_{\rm in} \over 4\pi R_c^3} {3-2\sqrt{(R/R_c)} \over (R/R_c)^2}~{\rm if}~0.2R_c \le R \le R_c
\en
and
\be
Q_{{\rm infall},a} = 0~{\rm if}~R < 0.2R_c~{\rm or}~R > R_c.
\en

As accretion proceeds, the central stellar mass increases and the disk gravitational
potential energy will become more negative.  In response, even in the absence of viscosity
disk material will move inward, implying additional accretion luminosity.  
The heating by this effect is
\be
Q_{{\rm grav},i} = {G\dot{M_*}\Sigma_i \over 2R},
\en
where $\dot{M_*}$ is change in the mass of the central star. 

The irradiation flux can be written as
\be
\label{eqn:heat_ext}
{\sigma T_{\rm ext}^4} = {f_* L_* \over 4\pi R^2} + {f_{\rm acc} L_{\rm acc} \over 4\pi R^2} + \sigma T_{\rm env}^4,
\en
where $L_*$ and $L_{\rm acc}$ are the stellar luminosity and the accretion luminosity, respectively, and $T_{\rm env}$ is the envelope cloud temperature. 
The coefficients $f_*$ and $f_{\rm acc}$ account for the non-normal irradiation of the disk
surface.  For the stellar irradiation we use $f_* = 0.1$ as in 
\citet{z10a} and assume the stellar luminosity
follows the mass-luminosity relation 
\be
\log \left ({L_* \over \lsun} \right) = 0.20 + 1.74\log \left( {M_* \over \msun} \right)
\en
which is an approximate power-law fit to pre-main sequence stars in the Taurus molecular cloud,
using the luminosities and effective temperatures from Kenyon
\& Hartmann (1995), and adopting the Siess et al. (2000) evolutionary tracks to obtain the masses.
The mass-luminosity relation is slightly modified from \citet{z10b}.

The inclusion of external heating by inner disk accretion is another new feature of
our calculations.  The accretion luminosity is calculated as 
\be
L_{\rm acc} = {GM_* \dot{M}\over 2 R_{\odot}},
\en
where we assume a typical T Tauri stellar radius.  In this case the appropriate value of
$f_{acc}$ is quite uncertain.  At low to moderate accretion rates, magnetospheric accretion
onto the star can occur at high latitudes, so that adoption of $f_{acc} = 0.1$, similar to
that used for the stellar photospheric irradiation, seems reasonable.  On the other hand,
at high accretion rates, the spectra of FU Ori objects provide no indication of magnetospheric
accretion \citep{hk96}, and irradiation of the outer disk by a relatively flat inner disk
should be much less effective \citep{bell99}.  However, if a substantial infalling envelope
surrounds the disk, it can capture much of the accretion luminosity and reradiate a significant
part toward the disk \citep{natta93}.  We therefore use both $f_{\rm acc} = 0.1$ and 0.01 
to examine the importance of this heating.

The last term in Equation (\ref{eqn:heat_ext}) is the flux from the envelope cloud whose temperature is assumed to 20~K.
Thus, the stellar luminosity irradiation $Q_*$ and the accretion luminosity irradiation $Q_{\rm acc}$ on the active layer become
\be
Q_* = {16 \over 3} {f_* L_* \over 4\pi R^2} {\tau_a \over 1+\tau_a^2}
\en
and
\be
Q_{\rm acc} = {16 \over 3} {f_{\rm acc} L_{\rm acc} \over 4\pi R^2} {\tau_a \over 1+\tau_a^2}.
\en
We note that the accretion luminosity irradiation $Q_{\rm acc}$ should be distinguished from the local viscous accretion heating $Q_{\rm vis}$.
The relative importance of the individual
heating terms, and their effects, during disk evolution will be discussed in \S\ref{sec:results}.

\subsection{Disk Viscosity}

The viscosity parameter $\alpha_i$ is the sum of the MRI viscosity parameter 
$\alpha_{M,i}$ and the GI viscosity parameter $\alpha_{Q,i}$.
The MRI viscosity parameter is assumed to have a fixed value of $\alpha_{MRI}$ only 
if a region can sustain the MRI.
Thus, the active layer viscosity parameter is always set to $\alpha_{M,a} = \alpha_{MRI}$ 
while the dead zone has MRI viscosity only if the midplane temperature is higher than a 
critical temperature $T_{\rm MRI}$ to produce sufficient ionization levels.
For the dead zone, we consider a residual viscosity as well as MRI viscosity and GI viscosity, 
$\alpha_d = \alpha_{M,d} + \alpha_{Q,d} + \alpha_{{\rm rd}}$.

The idea of the dead zone residual viscosity (DZRV) $\alpha_{\rm rd}$ is 
based on recent numerical magnetohydrodynamic simulations suggesting that
magnetic turbulence in the active layers can drive hydrodynamic turbulence
in the dead zone, implying a non-zero residual viscosity parameter 
$\sim10^{-3} - 10^{-5}$ (\citealt{bai11,okuzumi11,gressel12}).
Thus, for non-zero DZRV model we set
\be
\label{eqn:alpha_rd}
\alpha_{{\rm rd}} = {\rm min} \left(10^{-4},~f_{\rm rd} \alpha_{MRI}{\Sigma_a \over \Sigma_d} \right),
\en
where $f_{\rm rd}$ is the efficiency of accretion in the dead zone whose value is chosen 
to be $\leq 1$; this is intended to limit the effect of the active-layer induced
turbulence such that the mass accretion rate of the dead zone 
(approximately) does not exceed that of the active layer ($\dot{M}_d \le \dot{M}_a$).
This seems intuitively reasonable. We consider the upper limit
$f_{\rm rd} = 1$ and consider a case with $f_{\rm rd}=0.1$ as it is unlikely that
the active layer can be that effective in driving accretion.

Finally, the GI viscosity parameter is the same as in \citet{z10a},
\be
\alpha_{Q,i} = e^{-Q^2},
\en
where $Q$ is the Toomre parameter.

\section{Results }
\label{sec:results}

\subsection{Initial conditions}

We start with a $0.1~M_{\odot}$ central protostar surrounded by an $M_c = 1~M_{\odot}$ cloud.
We parameterize the cloud rotation in terms of
$\omega = \Omega_c / \Omega_{\rm b}$, where $\Omega_c$ is the (constant) angular frequency 
of the initial cloud, and $\Omega_{\rm b} = 2^{3/2} c_s^3/G M_{\rm c}$ 
is the breakup angular frequency at the outer cloud edge, and $c_s$ is the
(uniform) cloud sound speed.
Our fiducial models assume $\omega = 0.03$, 
which results in $\sim 15~\%$ larger cloud angular frequency than that 
used in the fiducial model of \citet{z10b} 
($\Omega_c \sim 1.15\times10^{-14}~{\rm rad~s^{-1}}$ in our model).
We set the maximum non-thermally ionized surface density $\Sigma_A$ to $100~{\rm g~cm^{-2}}$ for our fiducial choice and assume it is constant.
We assume $T_{\rm MRI} =$~1500~K and $\alpha_{\rm MRI} = 0.01$ for all calculations.
We adopt a cloud envelope temperature of $T_{\rm env}=20$~K, 
which yields a constant infall rate of $\sim3.4\times10^{-6}~\msunyr$.
This is $20~\%$ smaller than the infall rate for
conventional singular isothermal collapse model \citep{shu77} because of our modified 
infall model (see \S\ref{sec:density}).
The infall lasts for $\sim 0.24$~Myr, adding $0.8~M_{\odot}$ to the central star + disk in total.

\subsection{Zero dead zone viscosity model}

\citet{z10a} found that GI moves matter from the outer disk to the inner disk, 
leading to a pileup at $R \sim 2$~AU because GI is increasingly ineffective at small radii.  
Eventually enough material piles up to trap thermal energy that makes $T_d> T_{MRI}$, turns on the MRI thermally in the dead zone, and thus produces an outburst of accretion.
\citet{z10b} investigated the long-term evolution of such disks and found that 
the evolution can be divided into three stages.
The evolution starts with a quasi-steady disk accretion since the infall is to small radii where the inner disk can become hot enough to sustain the MRI thermally.
Then, it turns into the outburst stage as the infall occurs at radii $> 1$~AU.
After infall stops, the disk enters the T Tauri phase, having only a few GI-driven outside-in outbursts with a low mass accretion rate in between bursts.

\begin{figure}
\begin{center}
\includegraphics[scale=1]{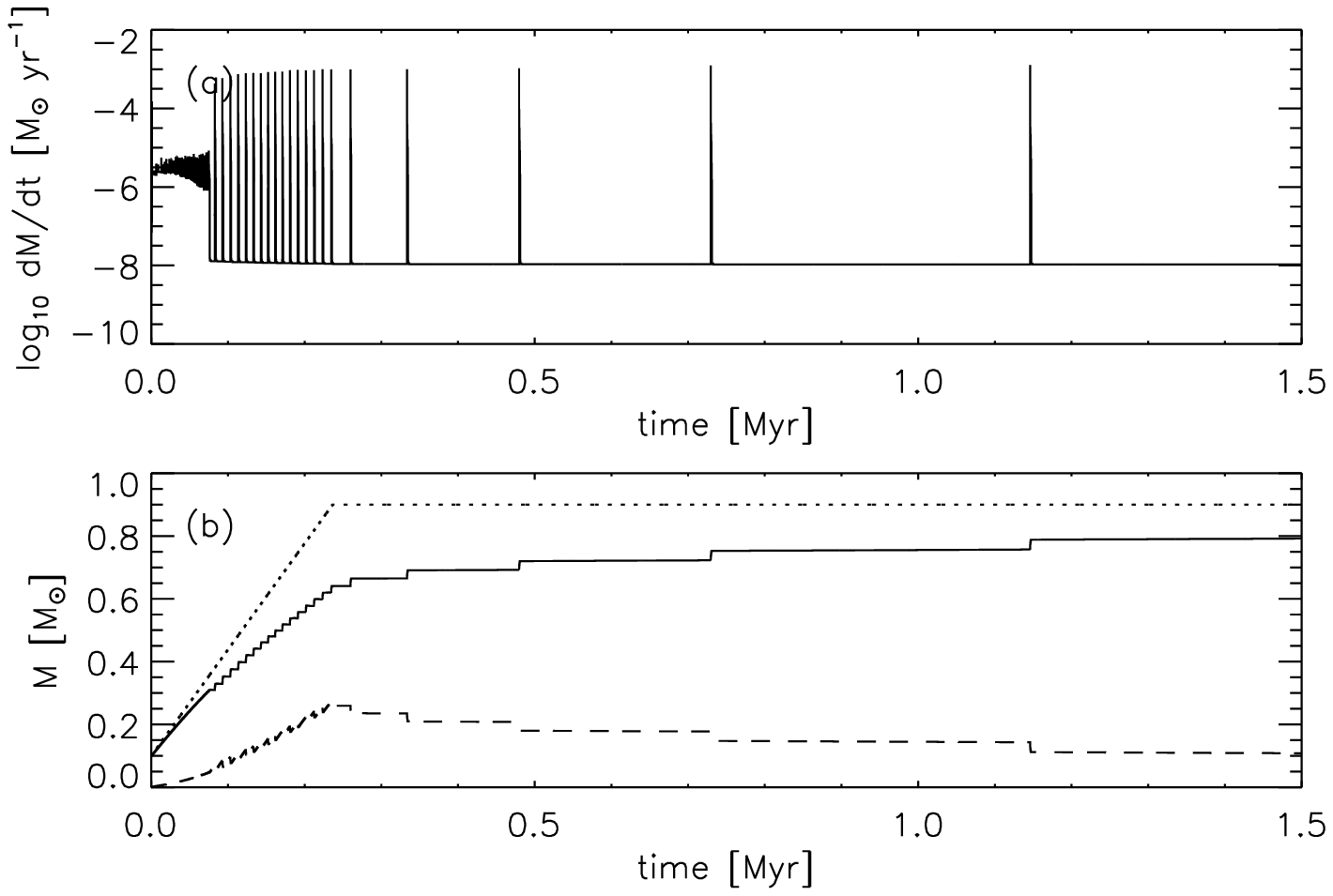}
\caption{(a) Mass accretion rate and (b) mass of the central star + disk (dotted curve), mass of the central star (solid curve), and mass of the disk (dashed curve) with time for the standard zero DZRV model.}
\label{fig:evol_z}
\end{center}
\end{figure}

Figure \ref{fig:evol_z} shows the mass accretion rate and the mass of the central star, the disk, and the central star + disk as a function of time.
The modified infall model and additional heating sources discussed in \S 2 
produce no qualitative difference in the overall evolution from \citet{z10b}.

\begin{figure}
\begin{center}
\includegraphics[scale=1]{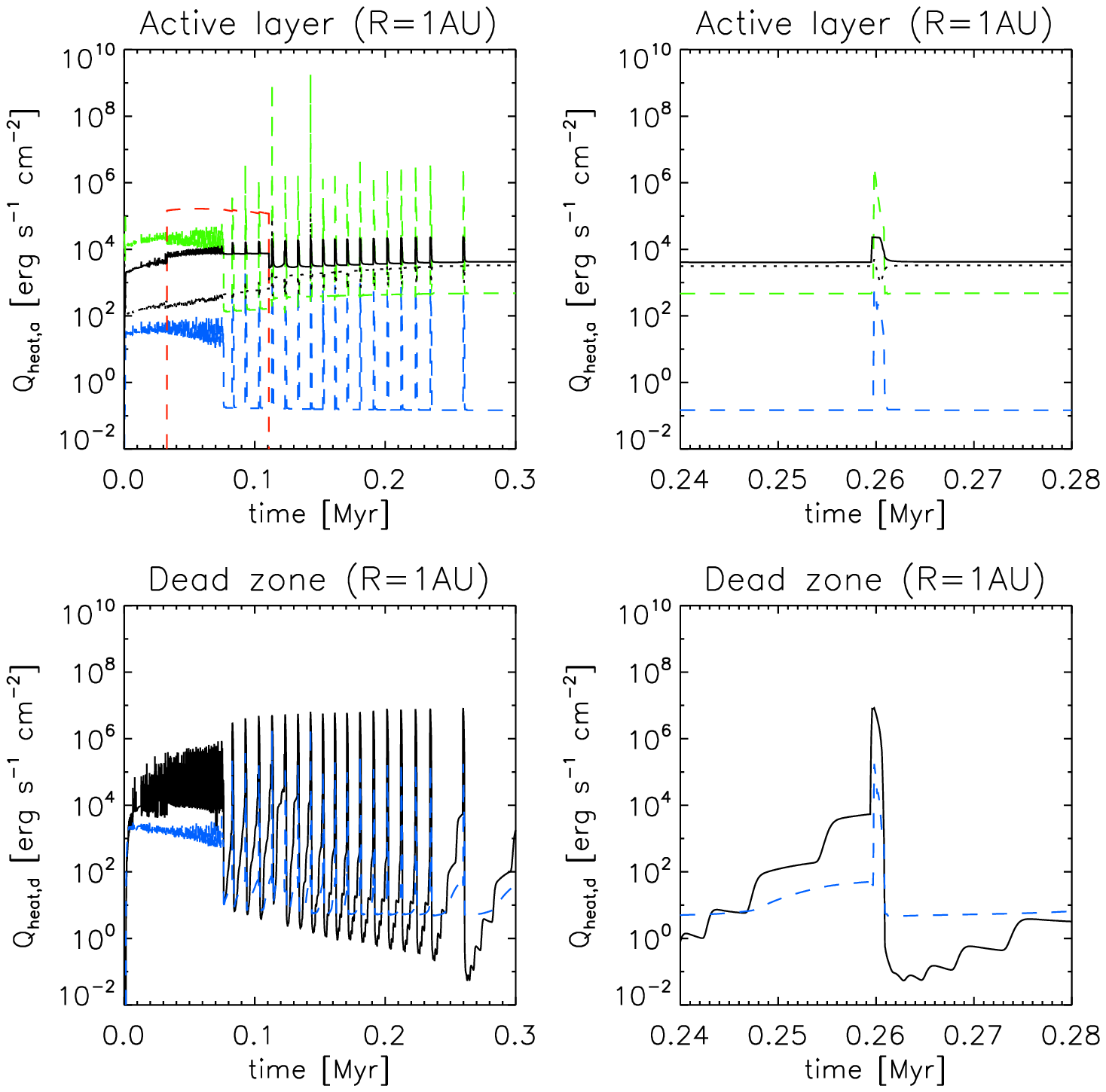}
\caption{Various heating sources of the active layer (upper panels) and the dead zone (lower panels) at $R=1$~AU for the first 0.3~Myr (left panels) and during a single outburst (right panels) with zero DZRV. Newly added heating sources in this paper are plotted in color while the heating sources have considered in \citep{z10a, z10b} are presented with black curves. Upper panels: $Q_{{\rm vis},a}$ (black solid), $Q_*$ (black dotted), $Q_{\rm infall}$ (red dashed), $Q_{\rm acc}$(green dashed), and $Q_{{\rm grav},a}$ (blue dashed) are presented. Lower panels: $Q_{{\rm vis},d}$ (black solid) and $Q_{{\rm grav},d}$ (blue dashed) are presented.}
\label{fig:heat_z1}
\end{center}
\end{figure}

\begin{figure}
\begin{center}
\includegraphics[scale=1]{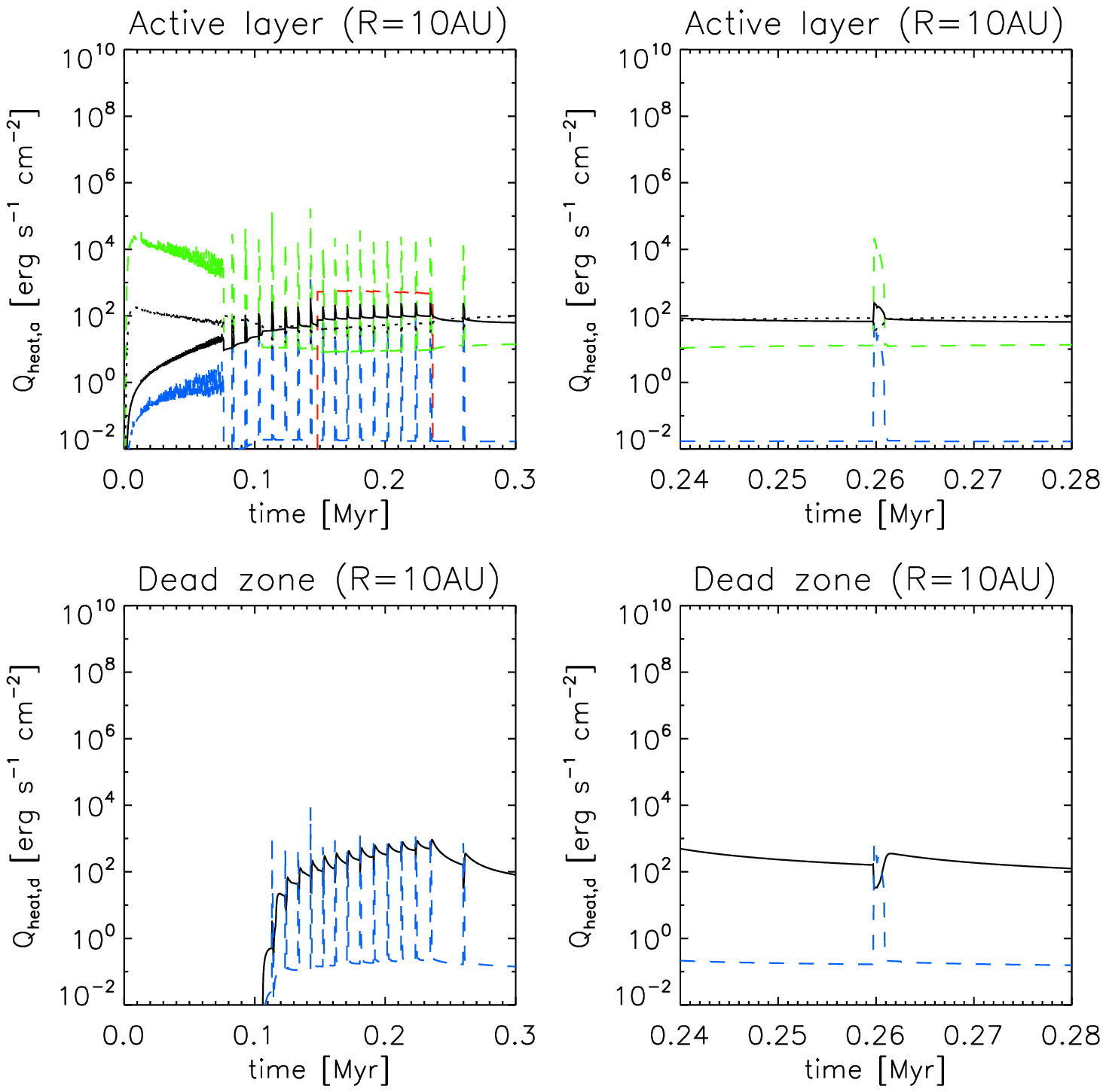}
\caption{Same as Figure \ref{fig:heat_z1} but at $R=10$~AU.}
\label{fig:heat_z10}
\end{center}
\end{figure}

\begin{figure}
\begin{center}
\includegraphics[scale=0.9]{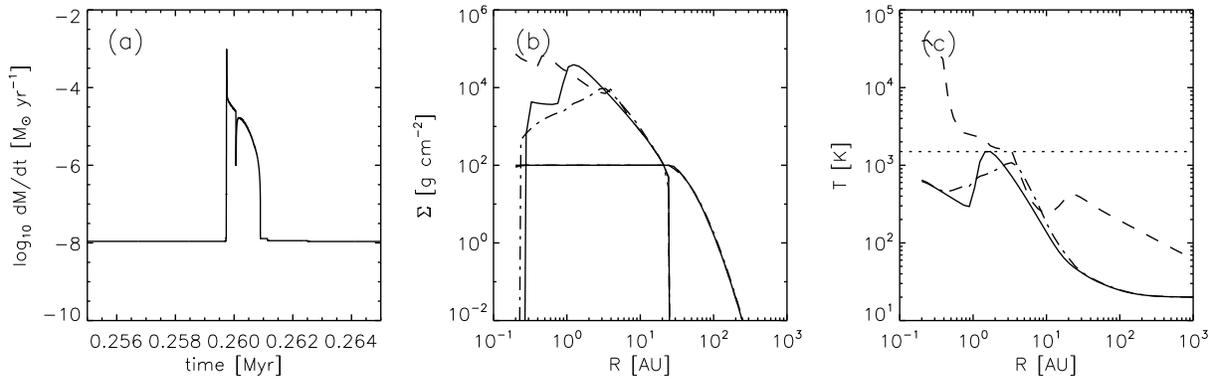}
\caption{(a) Mass accretion rate of a single outburst of the standard zero DZRV model. The ``drop out''
in accretion during the middle of the outburst is an artifact of the one-dimensional treatment \citep{z10a}. 
Radial profiles of (b) the surface densities and (c) the midplane temperatures before the outburst (solid curves), 
at the maximum accretion rate (dashed curves), and at the end (dash-dotted curves) of the burst are plotted.  
In panel (b), the curves with higher surface densities at the inner disk represent the dead zone surface density while the lower ones extend further out represent the active layer.
In panel (c), the dotted horizontal line represents the MRI activation temperature $T_{\rm MRI}=1500$~K.}
\label{fig:snap_z}
\end{center}
\end{figure}

The new heating sources for the first 0.3~Myr of the evolution at $R=1$ and 10~AU are 
presented in Figures \ref{fig:heat_z1} and \ref{fig:heat_z10}, 
together with the previously considered sources.
In these figures, newly added terms in this paper are plotted in color while other terms have considered in
\citet{z10a} and drawn in black.
As one can see, heating by the change in gravitational energy is usually several orders of magnitude smaller than other 
terms so that it makes no change in the evolution.
Infall heating provides comparable amount of heat to the active layer but is only limited to the region material falls onto,
increasing the local disk temperature slightly.
However, the large increase in the accretion luminosity during outburst produces enough
irradiation to make the outer disk temperature increase dramatically.
This is shown in Figure \ref{fig:snap_z}, where we show the mass accretion rate during a single outburst and 
the radial profiles of the disk surface density and the midplane temperature before outburst, at the maximum accretion rate, 
and at the end of the outburst.   
The temperature increase at the outer disk during outbursts does not affect the 
long-term evolution,
because the viscous time of the outer regions
($\sim 10^5$~yr) is much longer than the outburst timescale of $\sim 10^3$~yr.  

We also call attention to the jump in temperature at radii $< 0.5$~AU, which is due to thermal instability
\citep{z10a}.  This increase in temperature, which affects the behavior of the outburst of accreting material
onto the central star, would not have been found if we had taken an inner 
radius of $\gtrsim 1$~AU (see discussion in \S4).

The GI-driven outside-in bursts accrete $\sim0.027~M_{\odot}$ of material onto the central star and 
last for $\sim1340$ years on average.
Although the duration obtained in our calculation is longer than the typical outburst timescale
seen in  FU Ori, the timescale of an outburst can be scaled with a choice of 
$\alpha_{\rm MRI}$ since $\Delta t_{\rm burst} \sim R^2/\nu\propto\ {\alpha_{\rm MRI}}^{-1}$ (see \citealt{z10a}).
The outside-in bursts are rare, because the disk needs a lot of material to trigger MRI through GI while it is difficult to do so with zero DZRV.
The duty cycle of this model is thus pretty small, $\sim0.06$ during outburst stage and $\sim0.005$ during T Tauri phase.
The details of disk properties at 0.24 and 1~Myr and of outside-in bursts are summarized in Table {\ref{table:results}}.
All the outburst quantities in the table are time-averaged values after the initial quasi-accretion phase while those vary with time.

\subsection{Non-zero dead zone residual viscosity model}

\begin{figure}
\begin{center}
\includegraphics[scale=1]{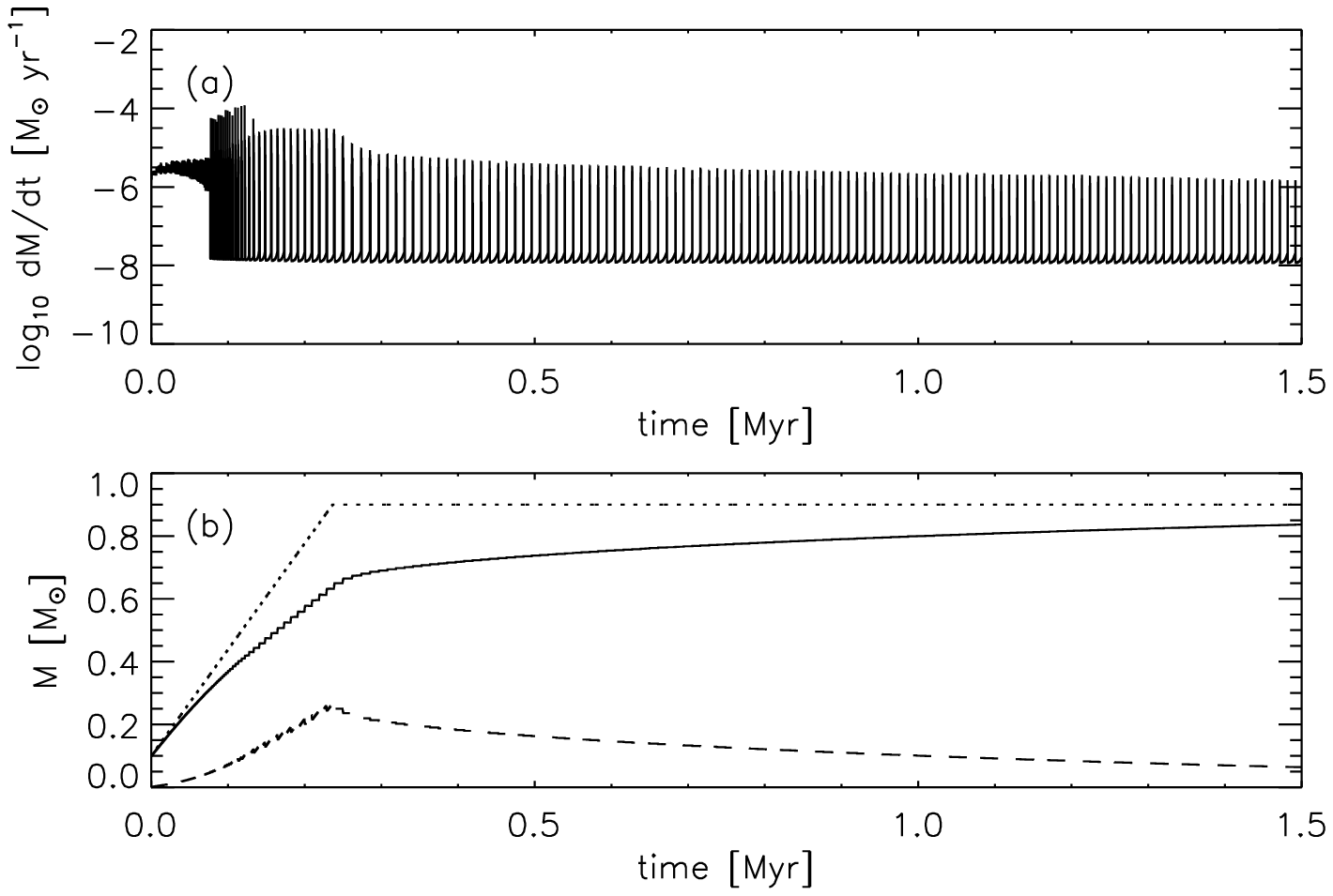}
\caption{(a) Mass accretion rate and (b) mass of the central star + disk (dotted curve), mass of the central star (solid curve), and mass of the disk (dashed curve) with time for the standard non-zero DZRV model.}
\label{fig:evol_nz}
\end{center}
\end{figure}

While the modified infall model and additional heating sources with zero DZRV make no qualitative change in the overall evolution, a finite DZRV makes a lot of difference in the long-term evolution and in the single outburst behavior as well.
Figure \ref{fig:evol_nz} shows the mass accretion rate and the mass of the central star, the disk, and the central star + disk as a function of time.
During infall the disk has a quasi-steady accretion phase at the beginning ($t \lesssim 0.08$~Myr) and the outburst stage follows, as in the zero DZRV case.
However, the evolution after the accretion phase is different in that the non-zero DZRV model shows a lot of smaller outbursts instead of a few large outbursts.

\begin{figure}
\begin{center}
\includegraphics[scale=1]{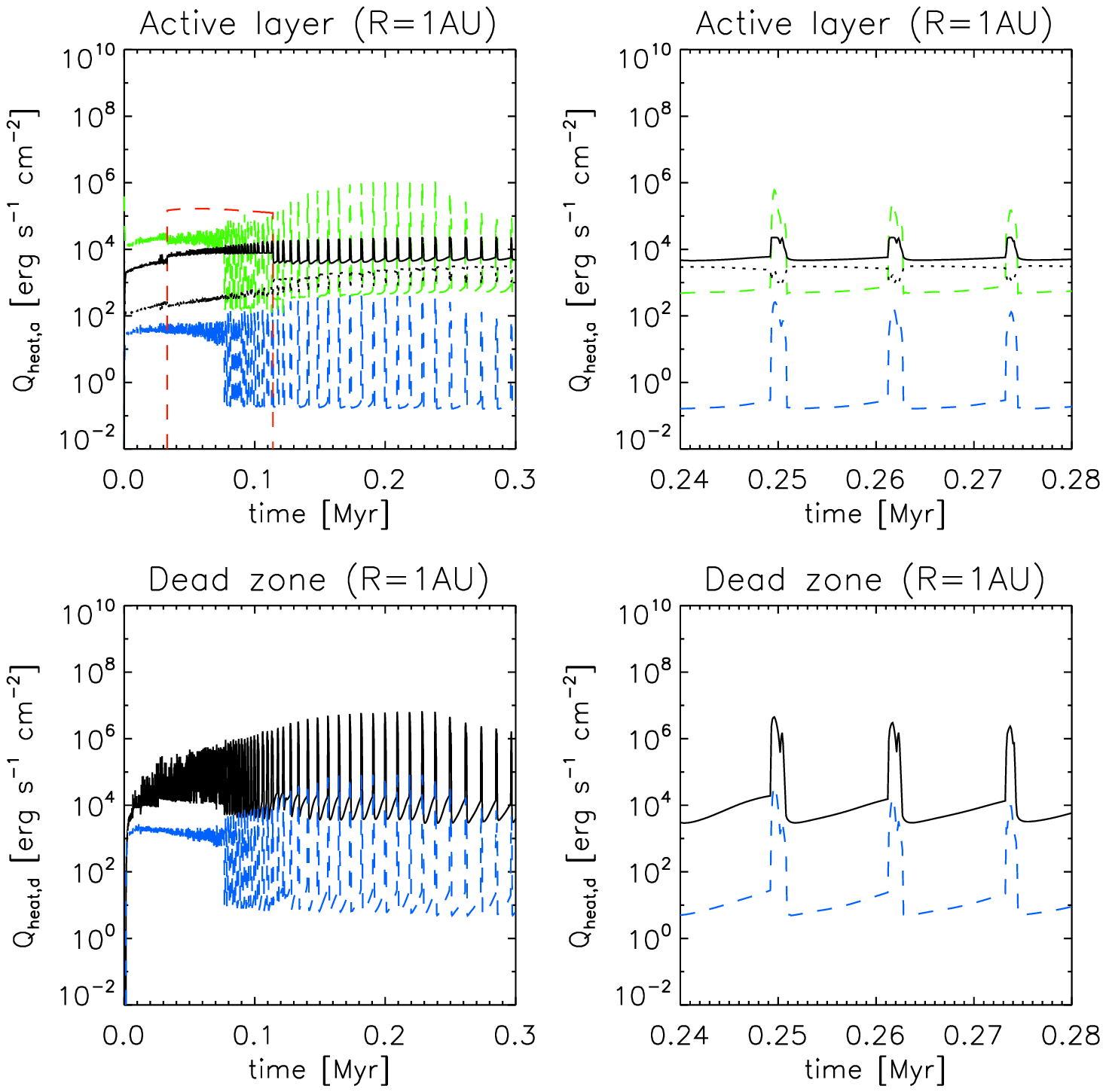}
\caption{Various heating sources of the active layer (upper panels) and the dead zone (lower panels) at $R=1$~AU for the first 0.3~Myr (left panels) and during a single outburst (right panels) with non-zero DZRV. Newly added heating sources in this paper are plotted in color while the heating sources have considered in \citep{z10a, z10b} are presented with black curves. Upper panels: $Q_{{\rm vis},a}$ (black solid), $Q_*$ (black dotted), $Q_{\rm infall}$ (red dashed), $Q_{\rm acc}$(green dashed), and $Q_{{\rm grav},a}$ (blue dashed) are presented. Lower panels: $Q_{{\rm vis},d}$ (black solid) and $Q_{{\rm grav},d}$ (blue dashed) are presented.}
\label{fig:heat_nz1}
\end{center}
\end{figure}

\begin{figure}
\begin{center}
\includegraphics[scale=1]{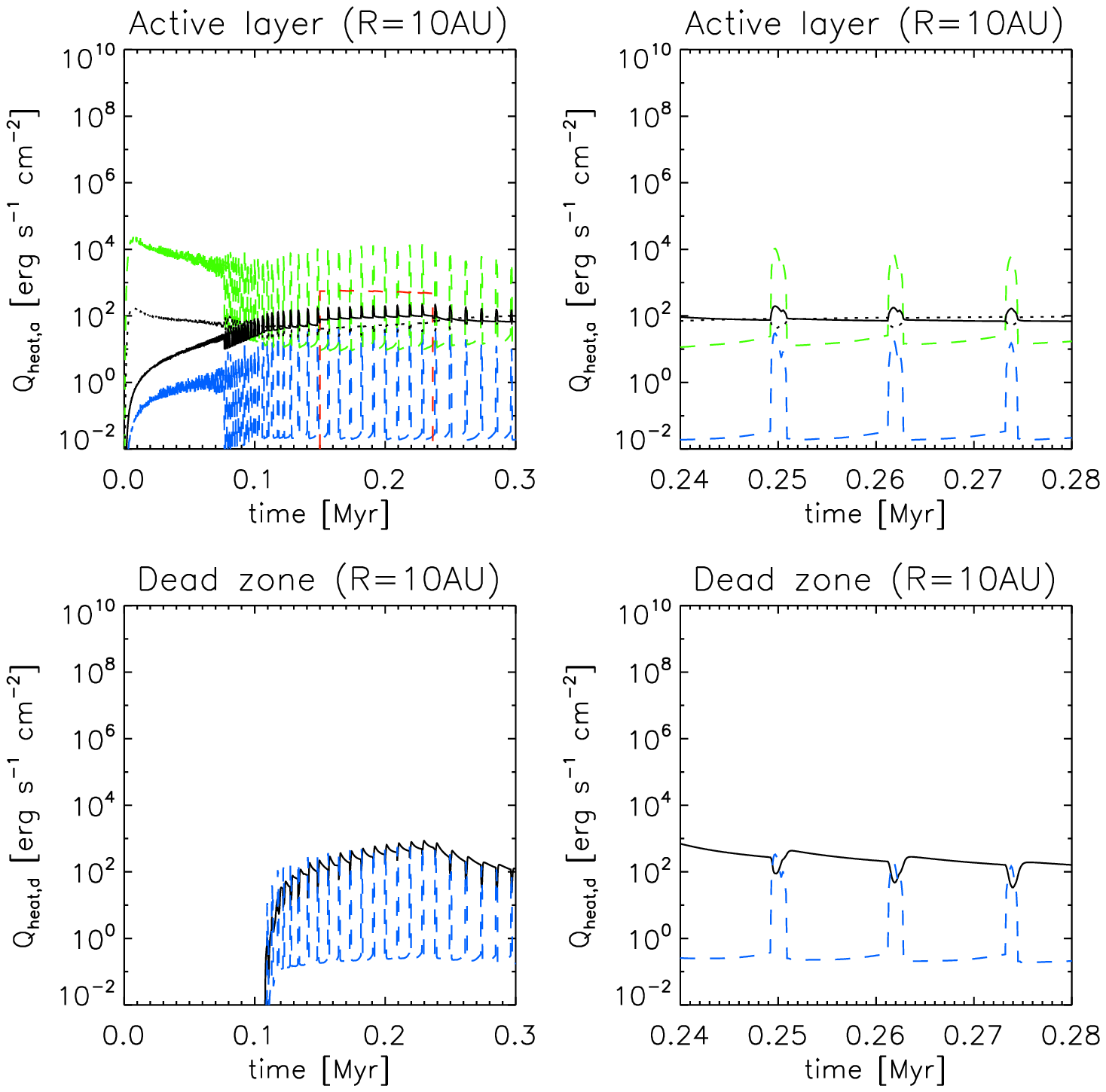}
\caption{Same as Figure \ref{fig:heat_nz1} but at $R=10$~AU.}
\label{fig:heat_nz10}
\end{center}
\end{figure}

Figure \ref{fig:heat_nz1} and \ref{fig:heat_nz10} show the heating sources of active layer and dead zone of the non-zero DZRV model at $R=1$ and 10~AU, respectively.
The major difference between this model and the zero DZRV model is that dead zone viscous heating provides a significant amount of heat at the inner disk even after infall ends.
We have run calculations with different $T_{\rm MRI}$ (1300~K and 1800~K) and found that the overall features are not sensitive to the choice of $T_{\rm MRI}$.
At the outer disk, the accretion luminosity irradiation is still important during bursts while the temperature increase is not as dramatic as in the zero DZRV model due to lower accretion peak.
Infall and gravitational heating are less important than others.

\begin{figure}
\begin{center}
\includegraphics[scale=0.9]{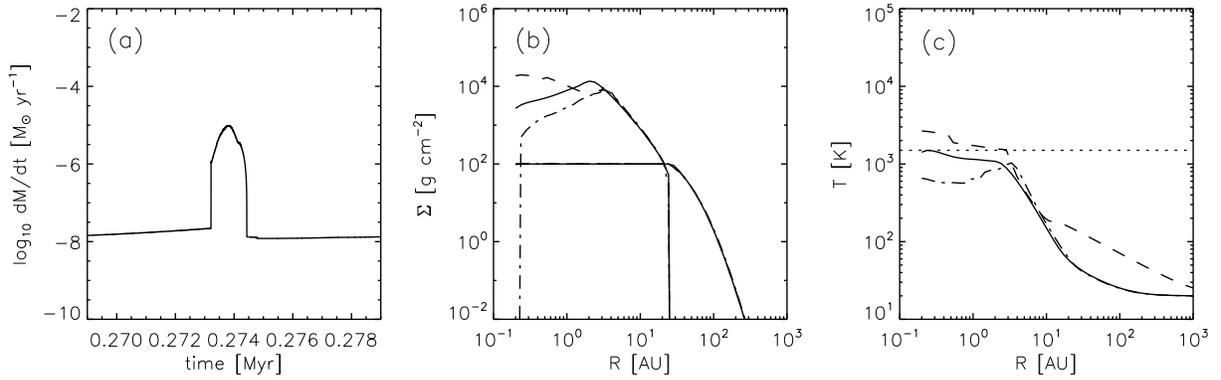}
\caption{(a) Mass accretion rate of a single outburst of the standard non-zero DZRV model. Radial profiles of (b) the surface densities and (c) the midplane temperatures at the beginning (solid curves), at the maximum accretion rate (dashed curves), and at the end (dash-dotted curves) of the burst. 
In panel (b), the curves with higher surface densities at the inner disk represent the dead zone surface density while the lower ones extend further out represent the active layer.
In panel (c), the dotted horizontal line represents the MRI activation temperature $T_{\rm MRI}=1500$~K.}
\label{fig:snap_nz}
\end{center}
\end{figure}

Figure \ref{fig:snap_nz}(a) shows the mass accretion rate during a single outburst with non-zero DZRV.
Radial surface density and midplane temperature profiles at the beginning, at the maximum accretion rate, and at the end of the burst are presented in Figure \ref{fig:snap_nz}(b) and (c).
The accretion behavior during a single outburst is remarkably different from that of the zero DZRV model.
The outburst has a peak of $\dot{M}_{\rm max}\sim10^{-5}~\msunyr$, which is about two orders of magnitude smaller than that of the outside-in bursts.
In addition, the accretion rate initially shows a rapid increase but has a slow rise time to its peak and a slow decrease after the peak as well.
In this model, the dead zone is able to transport material with the help of the non-zero DZRV.
Thus, the inner disk can be heated viscously and outbursts are initiated at the inner boundary 
of the disk before the GI piles up enough material at the middle of the disk ($R \sim 2$~AU) 
to initiate the MRI, which is the case for outside-in bursts.
The ionization front propagates out to several AU from the inner boundary. 
Since the inside-out bursts have an MRI active inner boundary from their initiation, the disk continues to dump material from its innermost part during the whole bursts.
Therefore, the system is not able to show a huge accretion rate as seen in outside-in bursts, but only generates moderate accretion rate.  Note again that the outburst triggers first at small radii, inside of $\sim 0.5$~AU (\S 4).

On average, the mass accreted onto the central star during a single inside-out burst is $\sim1.5\times10^{-3}~\msun$ and it lasts $\sim450$ years.
The inside-out bursts occur frequently enough to get a duty cycle of $\sim0.16$ during outburst stage and $\sim0.06$ during T Tauri phase.
The disk properties and outburst details of the non-zero DZRV model are summarized in Table \ref{table:results}.

\subsection{Efficiency of accretion luminosity irradiation}

As shown in the previous sections, irradiation by the inner disk
plays an important role during outburst on the temperature profile at the outer disk.
However, the efficiency of the accretion luminosity irradiation is uncertain as far as the non-normal irradiation of the disk is considered.
We thus test the effect of changing $f_{\rm acc}$ to 0.01, which is ten times smaller than the fiducial value.

We found essentially no change in the overall evolution of both zero and non-zero DZRV cases, since the accretion luminosity irradiation is several orders of magnitude smaller than main heating sources - active layer viscous heating and stellar irradiation - during the quiescent phase.
During outbursts, however, the accretion luminosity irradiation still dominates the heating even with a ten times smaller efficiency, making a significant difference to the outer disk temperature.
Not surprisingly, the increase in outer disk temperature during bursts is smaller than 
the standard cases by a factor of $\sim2$.

\subsection{Accretion Efficiency in Dead Zone}

Intuitively, it seems unlikely that the turbulence generated by the active layers within the dead zone
can transport as much mass as the active layer \citep[e.g.,][]{hartmann06}.  
In our models this happens when $\alpha_{\rm rd} \gtrsim 10^{-4}$, where $\Sigma_{d} \gtrsim 10^5~{\rm g~cm^{-2}}$.
This could be an overestimate of the efficiency with which the MRI turbulence in the active layers drives
accretion in the dead zone.  We therefore adjust the dead zone accretion efficiency $f_{\rm rd}$ to 0.1 
so that dead zone only has an accretion rate of $\sim 10~\%$ of the active layer at most.

\begin{figure}
\begin{center}
\includegraphics[scale=1]{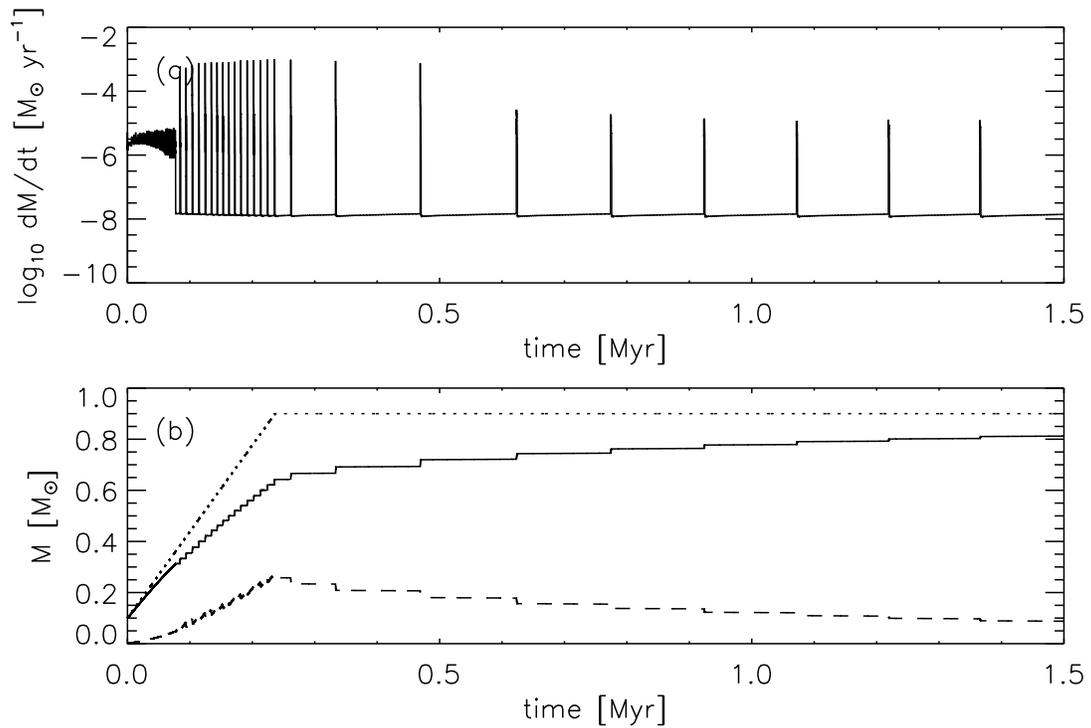}
\caption{(a) Mass accretion rate and (b) mass of the central star + disk (dotted curve), mass of the central star (solid curve), and mass of the disk (dashed curve) of non-zero DZRV model as a function of time, with $10~\%$ of accretion efficiency $f_{\rm rd}$ in the dead zone.}
\label{fig:evol_frd}
\end{center}
\end{figure}

Figure \ref{fig:evol_frd} shows the mass accretion rate and the mass of the central star, the disk, and the central star + disk as a function of time.
Initially, the evolution resembles that of the standard zero DZRV model rather than the non-zero DZRV model; the system shows a distinct outburst phase during infall.
This is because the mass that the dead zone can carry is now limited and thus generates less viscous heating at small radii than the standard non-zero DZRV case.
Therefore, mass piles up at large radii through the GI before inner disk gets heated and triggers inside-out bursts.
After infall ends, however, we still see inside-out bursts with much less frequency than the standard non-zero DZRV model, which is again due to less viscous heating at the inner disk.
The duty cycle during T Tauri phase of this model is only 0.015, which is four times smaller than that of the standard non-zero DZRV model.
First three outbursts after infall ends are outside-in bursts, since disk already collects enough material at outer disk during infall to make them.
This emphasizes importance of understanding the effect of MRI turbulence on dead zones (\S 4).

\subsection{Dependence on $\Sigma_A$ and $\alpha_{MRI}$}

While we use $\Sigma_A=100~{\rm g~cm}^{-2}$ as our fiducial value, several studies have pointed out that the active layer more likely has a lower surface density \citep[e.g.][]{sano00,bai09}. 
We thus test $\Sigma_A=20~{\rm g~cm}^{-2}$ and adjust the
MRI viscosity parameter $\alpha_{MRI}=0.05$ to maintain
roughly the same mass accretion rate ($\dot{M} \propto \alpha \Sigma$) during the quiescent phase as the standard cases (and also in agreement with typical T Tauri accretion rates).

\begin{figure}
\begin{center}
\includegraphics[scale=1]{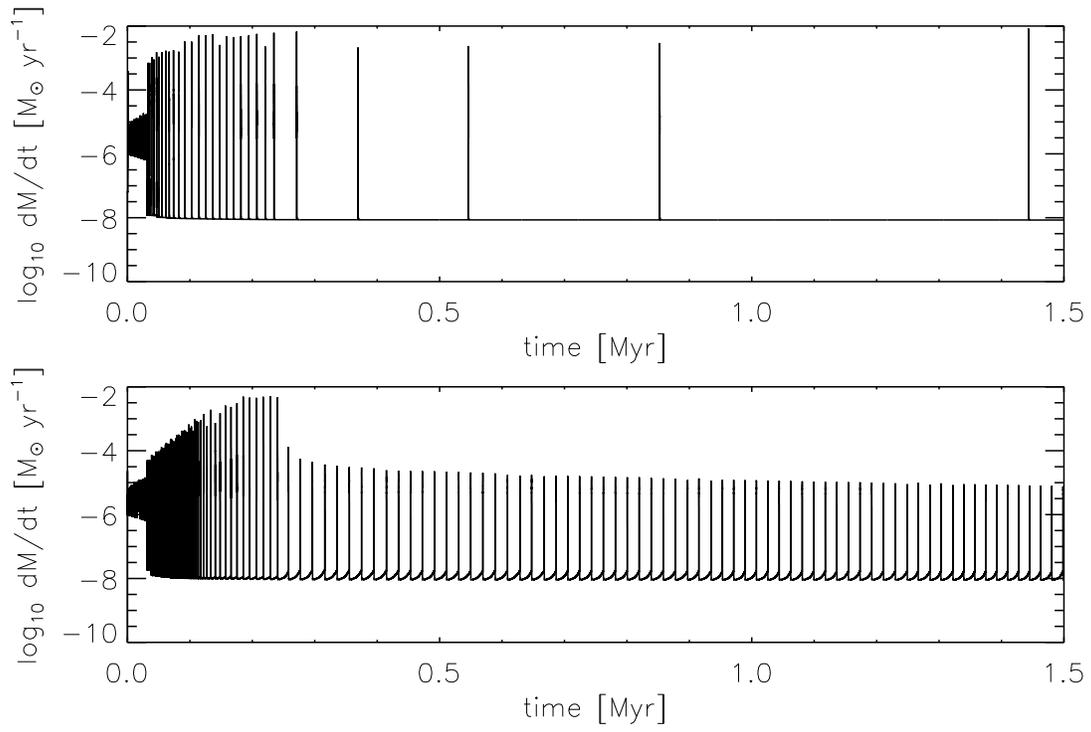}
\caption{Mass accretion rate of zero DZRV model (upper) and non-zero DZRV model (lower) as a function of time, with $\Sigma_A = 20~{\rm g~cm}^{-2}$ and $\alpha_{MRI}=0.05$.}
\label{fig:evol_alpha}
\end{center}
\end{figure}

\begin{figure}
\begin{center}
\includegraphics[scale=0.9]{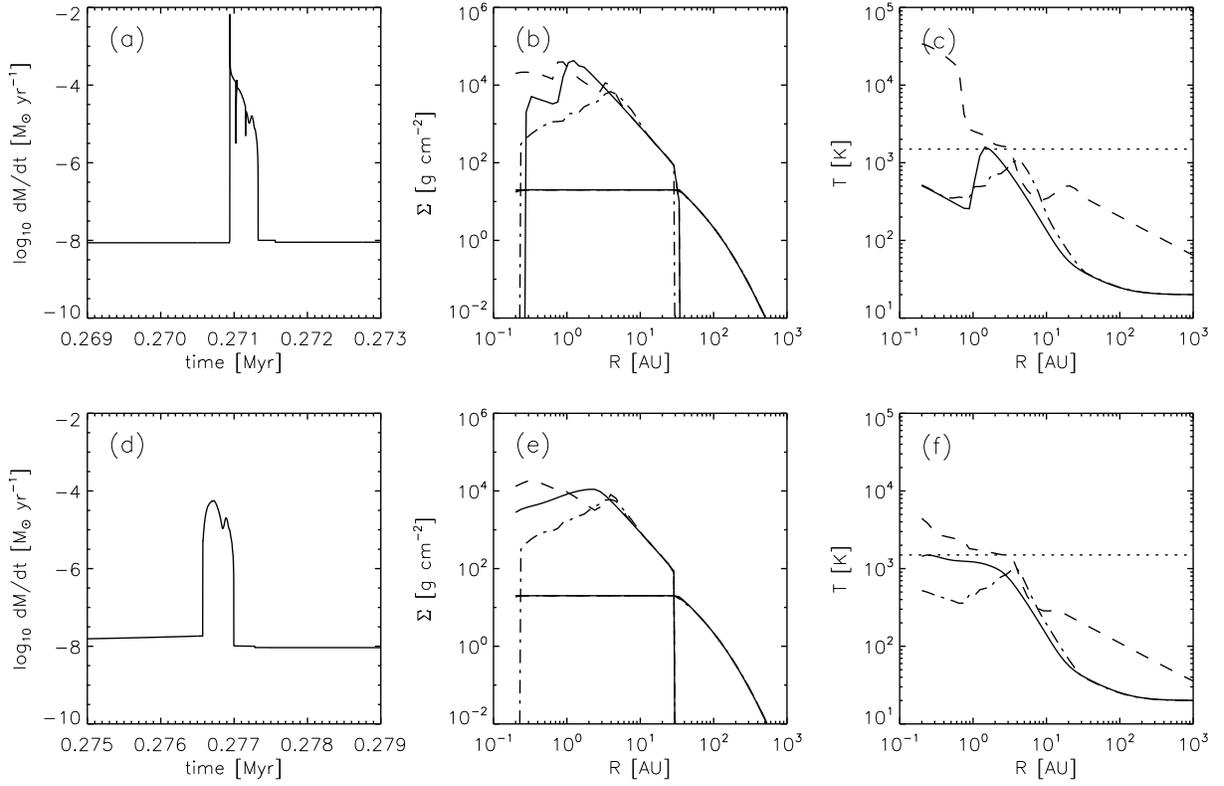}
\caption{Same as Figure \ref{fig:snap_z} (upper panels) and \ref{fig:snap_nz} (lower panels) but with $\Sigma_A = 20~{\rm g~cm}^{-2}$ and $\alpha_{MRI}=0.05$.}
\label{fig:snap_alpha}
\end{center}
\end{figure}

Figure \ref{fig:evol_alpha} shows the mass accretion rates of both zero and non-zero DZRV models adopting the lower value of $\Sigma_A$.
The mass accretion rate during a single outburst and radial surface density and midplane temperature profiles at the beginning, at the maximum accretion rate, and at the end of a single outburst of the both models are presented in Figure \ref{fig:snap_alpha}. 
Since the outburst timescale depends on the MRI viscosity parameter ($\Delta t_{\rm burst} \propto {\alpha_{MRI}}^{-1}$), the details of the outbursts, such as outburst duration and peak accretion rate, vary.
However, the overall evolution as well as the initiation of outbursts remain the same.
We see GI-induced MRI-driven outside-in bursts in the zero DZRV case and viscously triggered inside-out bursts in the non-zero DZRV case.
This is because the overall evolution and the initiation of outbursts are governed by the mass accretion during the quiescent phase, which we manage to be unchanged.
We note that the shorter timescales are in better agreement with FU Ori \citep[see][]{zhu07}.
The disk properties and outburst details are summarized in Table \ref{table:results}.

\section{Discussion}

Our simulations show that it is possible to obtain inside-out triggering of
accretion outbursts as well as outside-in bursts (e.g., Zhu \etal 2010a,b), 
with the former enhanced if there is finite dead zone residual viscosity.  
The two types of outbursts were
also obtained in the model developed by Bell \& Lin (1994; BL) for FU Ori outbursts.
In the BL model, the outbursts were due to thermal instability (TI), plus an assumed
increase in $\alpha$ from a very low value to a much higher value.  As Zhu \etal
(2007, 2008) showed, the TI model is inconsistent with observations
of FU Ori, because the high temperatures required limit the region of rapid
accretion to smaller radii than inferred from modeling the spectral energy distribution
including Spitzer Space Telescope data.  
Nevertheless, the finite dead zone residual viscosity
models are qualitatively similar to the basic feature of the BL models which
produce inside-out bursts; a small but finite viscosity allows material in the 
inner disk to produce enough trapping of viscously-generated heat to trigger 
a higher viscosity and eventually an outburst.  
As BL showed, such inner disk triggering leads to outbursts with slow rise times, 
qualitatively consistent with the observed outburst of V1515 Cyg (Herbig 1977; M. Ibrahimov, personal communication).

BL showed that outbursts with rapid rise times, such as observed in FU Ori and V1057 Cyg
\citep{herbig77}, required outside-in accretion events.  In the BL model, a large outer
perturbation of the disk was required.  In modern models, the event is triggered by
GI, which piles up material at larger radii than possible in the TI model \citep{armitage01,
vb06,vb07,vb09,vb10,z10a,z10b,martin12a,martin12b}.
Our current results build upon those of \citet{z10b} in that we clearly identify
some inside-out bursts during the main phase of infall (they were actually present
in the Zhu \etal simulation as well but were not emphasized).

Our results also bear similarities to outbursts in models of cataclysmic variables (CV).
Two different types of outbursts (outside-in and inside-out) in accretion disks around dwarf novae were first predicted by \citet{smak84}.
In CV models, mass transfer from the secondary rises the effective temperature of a disk annulus to $5000-8000$~K, which corresponds to hydrogen recombination inside the disk and thus triggers the TI \citep{menou99}.
While the dead zone residual viscosity parameter is the feature that changes outburst behavior in our model, CV models use the mass transfer rate to generate two different outbursts.
If the mass transfer rate is high, the accumulation timescale of transferred material is shorter than the viscous timescale, allowing material piles up at outer disk.
Thus, outside-in bursts are triggered.
In contrast, if mass transfer rate is low, the accumulation timescale becomes longer than the viscous timescale so that the outbursts outside-in outbursts are replaced by inside-out ones.
The resulting outburst behaviors of CV models are similar to ours; inside-out bursts have a smaller accretion peak and a slower rise time than outside-in ones (see Figures 3 and 4 of \citealt{hameury98}).

It is worth emphasizing that the outburst behavior of systems, observed at
optical and near-infrared wavelengths, is a result of accretion onto or near the
star, i.e.\ at radial scales $\lesssim 0.1$~AU.  While our models do not reach
magnetospheric or stellar radii, our inner boundary radius of $0.2$~AU is small enough
to capture behavior (thermal instability, inside-out bursts) which cannot be seen
in simulations with inner boundaries $> 1$~AU.  Thus, while our own treatment of
non-steady accretion has its limitations, time histories of accretion in
simulations with large inner disk radii must be treated with special caution.   
Using an inner boundary of a few AU, as in the series of papers by
\citet{vb06,vb07,vb09} or \citet{dunham12}, one would not find
the outburst behavior characterized by our models or those of \citet{armitage01} 
or \citet{martin12a,martin12b}.

Along these lines, we have found that the precise duration of the inside-out bursts during infall is sensitive
function of the value of the inner radius.  We do not explore this further here because there
are other major uncertainties in our treatment, such as the assumption of a constant
active layer surface density, 
the characterization of the GI via an $\alpha$ viscosity,
and the way in which we implement a finite dead zone viscosity.  
In the following paragraphs we discuss these issues in turn.

Advanced MHD treatments of the active layer are complex and involve
a number of unknowns, such as whether low-energy
cosmic rays can penetrate the accretion-driven winds, grain growth and settling,
the presence of metal ions, etc. (e.g., Sano \etal 2000; 
Ilgner \& Nelson 2006, 2008; Hirose \& Turner
2011; Perez-Becker \& Chiang 2011).  Martin \etal (2012a,b) argue that a large critical
Reynolds number is necessary for transport to occur, resulting ``active'' regions
very different than the constant $\Sigma_a$ we use.  However, Martin \etal (2012b)
predict essentially no accretion onto the central star in between outbursts,
whereas the pre-outburst spectrum of V1057 Cyg shows emission lines typical of
T Tauri stars accreting at $\sim 10^{-8} \msunyr$ (Herbig 1977).  
More recently, Miller \etal (2011) showed that the classical (accreting) T Tauri star
LkHa 188-G4 underwent an FU Ori-type eruption in 2009.  Finally, 
a pre-outburst spectrum of the FU Ori object V733 Cep (Reipurth \etal 2007) also showed
characteristic accreting T Tauri emission lines (B. Reipurth, personal communication).  
Thus the pre-outburst state of at least some FU Ori objects is one of accretion at
rates typical of T Tauri stars, which we obtain with our adopted value of 
$\alpha_a \Sigma_a$, at least in the inner disk.

Similarly, our treatment of the GI with an $\alpha$ viscosity is crude; as in
the case of the MRI, three-dimensional simulations are required to treat the GI
properly (e.g., Rice \etal 2003; Boley \etal 2006; Durisen
\etal 2007, and references therein).  Two-dimensional simulations also do a better
job of capturing the GI than our treatment (e.g., Vorobyov \& Basu 2006, 2007, 2009, 2012).
However, as pointed out by Zhu \etal (2009), and as shown in Zhu \etal (2010a,b) and
the present simulations, the GI becomes harder and harder to sustain as one moves
to smaller radii, whereas triggering of the MRI becomes easier.  

Finally, the presence and behavior of non-zero viscosity in dead zones is  
highly uncertain.  While simulations of resistive, stratified disks in shearing
boxes appear to show that the active layer produces a non-zero $\alpha$ in the
central ``dead'' layers (Fleming \& Stone 2003; Okuzumi \& Hirose 2011; 
Gressel, Nelson, \& Turner 2012), the level to which this occurs, the amount
of mass transport involved, and precisely where energy is dissipated is unclear.
For example, our model assumes that the wave energy is dissipated near the midplane,
and thus the heat generated can be trapped radiatively by the opacity of the
disk at the midplane; however, it is possible that dissipation is concentrated
at higher levels (N. Turner, personal communication).  

In all, these uncertainties show that our current results must be taken as suggestive
rather than predictive for variations of accretion in young stellar objects.
Nevertheless, these simple models, which can be evolved easily for significant
evolutionary timescales, illustrate the potential information on transport processes
in protostellar and protoplanetary disks that might ultimately be gleaned from
the observed accretion outbursts.  Due to the increasing monitoring of young stellar
objects, it is becoming increasingly clear that a wide variety
of accretion behavior is exhibited in young stars, emphasizing that further progress on 
challenging problems of globally simulating MRI and GI in protostellar disks
may pay rich dividends. 

\section{Summary}

In this paper, we have extended the one-dimensional, two-zone model of long-term protostellar disk evolution with infall, which is previously introduced in \citet{z10a,z10b}.
Our modified models include a revised treatment of infall, 
enhanced disk heating, and possible non-zero viscosity in the dead zone.
While the former two changes produce no qualitative difference in the overall evolution from the \citet{z10b}, we find that the presence of a small but finite dead zone viscosity can trigger inside-out bursts initiated at or near the inner edge of the disk through dead zone viscous heating, instead of GI-induced MRI-driven outside-in bursts with zero dead zone viscosity.
These inside-out bursts not only bear a qualitative resemblance to the outburst behavior of one FU Ori objects, V1515~Cyg, but emphasize a careful treatment of the inner disk regions in simulations.

Given the uncertainties, our results are rather suggestive than predictive.
However, two types of outbursts seen in FU Ori objects
can be successfully reproduced by the simple $\alpha$ treatment in the dead zone.
This difference in accretion behavior could be a 
potential probe of transport efficiency in the dead zone.

\acknowledgments

We acknowledge useful conversations with Bo Reipurth and Neal Turner.
This work was supported in part by NASA grant NNX08A139G and by the
University of Michigan.

%figure 1

%figure 2

%figure 3

%figure 4

%figure 5

%figure 6

%figure 7

%figure 8

%figure 9

%figure 10

%figure 11

%figure 12

\clearpage

\begin{deluxetable}{ccccccccccccc}
\tablecolumns{13}
\tabletypesize{\tiny}
\tablecaption{Parameters and results\label{table:results}}
\tablewidth{0pt}
\tablehead{
\colhead{$\alpha_{\rm rd}$} & 
\colhead{$\alpha_{MRI}$} & 
\colhead{$\Sigma_A$} & 
\colhead{$R_{\rm dead}$\tablenotemark{a}} &
\colhead{$M_*$\tablenotemark{a}} &
\colhead{$M_{\rm disk}$\tablenotemark{a}} &
\colhead{$M_{\rm in}$\tablenotemark{a}} &
\colhead{$M_{\rm out}$\tablenotemark{a}} &
\colhead{$M_{\rm burst}$\tablenotemark{b}} &
\colhead{$\dot{M}_{\rm max}$\tablenotemark{b}} &
\colhead{$\Delta t_{\rm{burst}}$\tablenotemark{b}} &
\colhead{$D_O$\tablenotemark{c}} &
\colhead{$D_T$\tablenotemark{d}} \\
\colhead{} & 
\colhead{} & 
\colhead{(${\rm g~cm}^{-2}$)} & 
\colhead{(AU)} &
\colhead{($M_{\odot}$)} &
\colhead{($M_{\odot}$)} &
\colhead{($M_{\odot}$)} &
\colhead{($M_{\odot}$)} &
\colhead{($M_{\odot}$)} &
\colhead{($M_{\odot}~{\rm yr}^{-1}$)} &
\colhead{(yr)} &
\colhead{} &
\colhead{}
 }
\startdata
zero & 0.01 & 100 & 29/5.5 & 0.64/0.76 & 0.26/0.14 & 0.21/0.08 & 0.05/0.06 & 0.027 & $1.11\times 10^{-3}$ & 1340 & 0.06 & 0.005  \\
non-zero & 0.01& 100 & 24/6.5 & 0.65/0.80 & 0.25/0.10 & 0.20/0.04 & 0.05/0.06 & $1.5\times10^{-3}$ & $3.17\times 10^{-6}$ & 450 & 0.16 & 0.06\\
zero & 0.05 & 20 & 33.9/7.6 & 0.65/0.80 & 0.25/0.10 & 0.22/0.08 & 0.03/0.02 & 0.028 & $4.34\times 10^{-3}$ & 310 & 0.016 & 0.002  \\
non-zero & 0.05 & 20 & 33.9/7.6 & 0.65/0.82 & 0.25/0.08 & 0.22/0.05 & 0.04/0.03 & $4.6\times10^{-3}$ & $4.26 \times 10^{-5}$ & 200 & 0.08 & 0.015
\enddata
\tablenotetext{a}{Quantities are taken at 0.24 and 1~Myr.}
\tablenotetext{b}{Outburst quantities are averaged over time after infall ends.}
\tablenotetext{c}{Duty cycle during outburst stage.}
\tablenotetext{d}{Duty cycle during T Tauri phase.}
\end{deluxetable}

\end{document}